\documentclass[conference]{IEEEtran}
\IEEEoverridecommandlockouts
\usepackage{cite}
\usepackage{amsmath,amssymb,amsfonts}
\usepackage{graphicx}
\usepackage{textcomp}
\usepackage{xcolor}
\usepackage{balance}
\usepackage{cite}
\usepackage{amsmath,amssymb,amsfonts}
\usepackage{amsmath}
\usepackage{amsthm}
\DeclareMathOperator*{\argmax}{argmax}
\DeclareMathOperator*{\argmin}{argmin}
\usepackage{graphicx}
\usepackage{textcomp}
\usepackage{xcolor}
\usepackage{graphicx}
\usepackage{float}
\usepackage{subfigure}
\usepackage{amsmath}
\usepackage{amsfonts,amssymb}
\usepackage{mathrsfs}
\usepackage{mathtools}
\usepackage{bm}
\usepackage{multirow}
\usepackage{array}
\usepackage{amssymb}
\usepackage{amsmath}
\usepackage{cite}
\usepackage{url}
\usepackage{xcolor}
\usepackage{cite,graphicx,amsmath,amssymb,cases}
\usepackage{subfigure}
\usepackage{fancyhdr}
\usepackage{mdwmath}
\usepackage{mdwtab}
\usepackage{caption}
\usepackage{amsthm}
\usepackage{setspace}
\usepackage{bm}
\usepackage{mathtools}
\usepackage{dsfont}
\usepackage{bbm}
\usepackage{algorithm}
\usepackage{algorithmic}
\newtheorem{remark}{Remark}
\newtheorem{theorem}{Theorem}

\newtheorem{lemma}{Lemma}

\newtheorem{corollary}{Corollary}

\makeatletter
\newcommand{\biggg}{\bBigg@{3}}
\newcommand{\Biggg}{\bBigg@{3.5}}
\makeatother
\def\BibTeX{{\rm B\kern-.05em{\sc i\kern-.025em b}\kern-.08em
    T\kern-.1667em\lower.7ex\hbox{E}\kern-.125emX}}
\begin{document}

\title{Secure Antenna Selection and Beamforming in MIMO Systems}

\author{\IEEEauthorblockN{Zhenqiao Cheng$^{\dag}$, Nanxi Li$^{\dag}$, Ruizhe Long$^{\sharp}$, Jianchi Zhu$^{\dag}$, Chongjun~Ouyang$^{\star\ddag}$, Peng Chen$^{\dag}$}
$^\dag$6G Research Centre, China Telecom Beijing Research Institute, Beijing, 102209, China\\
$^{\sharp}$University of Electronic Science and Technology of China, Chengdu, 611731, China\\
$^{\star}$School of Electronic Engineering and Computer Science, Queen Mary University of London\\
$^\ddag$School of Electrical and Electronic Engineering, University College Dublin\\
Email: $^{\dag}$\{chengzq, linanxi, zhujc, chenpeng11\}@chinatelecom.cn\\
$^{\sharp}$ruizhelong@gmail.com, $^{\star}$c.ouyang@qmul.ac.uk}
\maketitle

\begin{abstract}
This work proposes a novel joint design for multiuser multiple-input multiple-output wiretap channels. The base station exploits a switching network to connect a subset of its antennas to the available radio frequency chains. The switching network and transmit beamformers are jointly designed to maximize the weighted secrecy sum-rate for this setting. The principal design problem reduces to an NP-hard mixed-integer non-linear programming. We invoke the fractional programming technique and the penalty dual decomposition method to develop a tractable iterative algorithm that effectively approximates the optimal design. Our numerical investigations validate the effectiveness of the proposed algorithm and its superior performance compared with the benchmark.
\end{abstract}
\section{Introduction}
Information security has always been a critical issue for wireless communications due to the broadcast property of the wireless medium. This fact spawned the development and application of physical layer (PHY) security, where the transmitter exploits secrecy channel coding to ensure perfect security, i.e., eavesdroppers cannot decipher confidential information from wiretapped messages. A feasible method to further boost the secrecy performance at the PHY is to exploit the spatial degrees of freedom offered by multiple antennas \cite{Chen2017}. However, its fully digital implementation with a dedicated radio frequency (RF) chain at each antenna suffers from expensive hardware costs and excessive energy consumption. To address these challenges, numerous potential technologies have been introduced over the last few years. Among them is antenna selection, a technique to set only a small subset of antennas active in each coherence time \cite{Ouyang2020}. This technique can alleviate the requirement on the number of RF transceivers without significantly sacrificing the secrecy performance \cite{Asaad2018_JSAC}.

The past years have seen increasingly rapid advances in antenna selection algorithm design in multiple-input multiple-output (MIMO) wiretap channels. Most initial efforts focused on the secrecy performance achieved by single antenna selection; see \cite{Ouyang2020} and the references therein. Extension to multiple antenna selection settings with single legitimate and eavesdropping user terminals (UTs) were discussed in \cite{Asaad2018_JSAC, Tian2020, Ouyang_2019}. These works were further generalized to the multiuser case \cite{Bereyhi2021, Bereyhi2018}. Despite extending the basic studies on this topic, the available literature is still restricted to some particular scenarios. Specifically, the results in \cite{Bereyhi2021, Bereyhi2018} were based on the maximal ratio transmission (MRT) protocol, and a joint beamforming and antenna selection design is still lacking.
\subsection{Contributions}
This paper studies the secure transmission in a multiuser MIMO wiretap channel where the base station (BS) sets only a subset of available antennas active for communications. We maximize the secrecy throughput by proposing a joint design that optimizes the switching network and beamformers at the BS. The principal design problem belongs to mixed-integer non-linear programming (MINLP). We address this challenging problem through the following contributions: 1) We propose a penalty dual decomposition (PDD)-based method to tackle the non-convex joint design problem via capitalizing on the fractional programming (FP) technique. 2) To further reduce the computational complexity, we propose an alternative algorithm based on sequential optimization (SO). We show through numerical experiments that this algorithm involves reduced complexity at the expense of a minor secrecy performance loss. Our numerical results verify the capability of the proposed approaches to outperform the benchmark significantly.
\subsection{Notation}
Throughout this paper, scalars, vectors, and matrices are denoted by non-bold, bold lower-case, and bold upper-case letters, respectively. The notations $[\mathbf A]_{i,j}$ and $[\mathbf a]_{i}$ denote the $(i,j)$th entry and the $i$th entry of matrix $\mathbf A$ and vector $\mathbf a$, respectively. The identity matrix, zero matrix, and all-one vector are represented by $\mathbf I$, $\mathbf 0$, and $\mathbf 1$, respectively. The Hadamard product is shown by $\odot$ and $[K]$ represents the integer set $\{1, \ldots ,K\}$.
\section{System Model and Problem Formulation}
\begin{figure}[!t]
\centering
\includegraphics[height=0.2\textwidth]{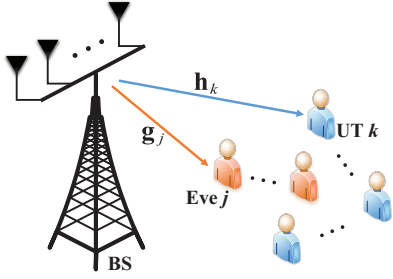}
\caption{Illustration of a multiuser MIMO wiretap channel}
\label{Figure1}
\end{figure}

Consider MIMO secure transmission where one $M$-antenna BS sends messages simultaneously to $K$ legitimate UTs in the same time/frequency resources, while these messages are confidential to $J$ adversaries that are treated as eavesdroppers (Eves), as depicted in {\figurename} {\ref{Figure1}}. We assume that each UT $k\in[K]$ and each Eve $j\in[J]$ are equipped with a single antenna for receiving. Let ${\mathbf h}_k\in{\mathbbmss C}^{M\times1}$ and ${\mathbf g}_j\in{\mathbbmss C}^{M\times1}$ denote the UT $k$-to-BS and Eve $j$-to-BS channel vectors, respectively. Furthermore, we denote ${x}_k\in{\mathbbmss{C}}$ as the securely coded data symbol dedicated to UT $k$, with zero mean and unit variance. The data symbol $x_k$ is independent of the symbols dedicated to other UTs, i.e., $\mathbbmss{E}\left\{x_kx_{k'}^{\mathsf{H}}\right\}={0}$ for $k\neq k'$.

The system operates in the time-division duplexing (TDD) mode, where the instantaneous channel state information (CSI) can be estimated via pilot sequences in the uplink training phase. This work focuses on passive eavesdropping, where the eavesdroppers are registered UTs in the system but distrusted by the legitimate UTs. In such a scenario, the eavesdroppers will participate in the uplink training phase; thus, the BS can estimate their CSI. We assume that the pilots are mutually orthogonal and that the estimation error is negligible. The BS thus learns perfectly the CSI $\{{\mathbf h}_k\}_{k=1}^{K}$ and $\{{\mathbf g}_j\}_{j=1}^{J}$.
\subsection{Transmit Antenna Selection}
The BS has $N<M$ RF chains and thus uses a switching network to select a subset of transmit antennas. This switching network connects the selected antennas to the available $N$ RF chains at the BS. As a result, the observations at UT $k$ and Eve $j$ are, respectively, given by
\begin{align}
&y_k={\mathbf h}_k^{\mathsf H}{\bm\Delta}\sum_{k=1}^{K}\mathbf{w}_kx_k+n_k,\\
&\overline{y}_j={\mathbf g}_j^{\mathsf H}{\bm\Delta}\sum_{k=1}^{K}\mathbf{w}_kx_k+\overline{n}_j,
\end{align}
where ${\mathbf w}_k\in{\mathbbmss C}^{M\times1}$ is the beamforming vector associated with symbol $x_k$, $n_k\sim{\mathcal{CN}}(0,\sigma_k^2)$
 and $\overline{n}_j\sim{\mathcal{CN}}(0,\delta_j^2)$ denote the additive white Gaussian noises (AWGNs) with $\sigma_k^2$ and $\delta_j^2$ being the noise powers. Moreover, ${\bm\Delta}={\mathsf{diag}}\{s_1,\ldots,s_M\}$ is the antenna selection matrix with $[s_1,\ldots,s_M]^{\mathsf T}\triangleq{\mathbf s}$ and
\begin{align}
s_m=\begin{cases}
1& {\text{antenna}}~m~{\text{is selected}}\\
0& {\text{otherwise}}
\end{cases}.
\end{align}
We note that ${\bm\Delta}{\bm\Delta}^{\mathsf H}={\bm\Delta}$. Assume that each UT $k$ and each Eve $j$ have access to instantaneous CSI of their own channels with properly designed pilot signals. The signal-to-interference-plus-noise ratio (SINR) at UT $k$ is given by
\begin{align}
\gamma_k=\frac{\lvert{\mathbf h}_k^{\mathsf H}{\bm\Delta}{\mathbf w}_k\rvert^2}{\sigma_k^2+\sum_{k'\neq k}\lvert{\mathbf h}_k^{\mathsf H}{\bm\Delta}{\mathbf w}_{k'}\rvert^2}.
\end{align}
\subsection{Performance Metric: Weighted Secrecy Sum-Rate}
From a worst-case design perspective, we assume that all eavesdroppers cooperate to overhear the secure transmission and can cancel out the interference of other legitimate UTs. Given a beamforming matrix ${\mathbf W}=[{\mathbf w}_1,\ldots,{\mathbf w}_K]$ and a selection matrix ${\bm\Delta}$, the maximum secrecy rate to UT $k\in[K]$ in the worst-case scenario is given by
\begin{align}\label{Secrecy_Rate_UT}
{\mathcal R}_{k}=[\log_2\left(1+\gamma_k\right)-\log_2\left(1+{\overline{\gamma}}_k\right),0]^{+},
\end{align}
where $[\cdot]^{+}\triangleq\max\{\cdot,0\}$ and ${\overline{\gamma}}_k=\sum_{j=1}^{J}\frac{1}{\delta_j^2}\lvert{\mathbf g}_j^{\mathsf H}{\bm\Delta}{\mathbf w}_k\rvert^2$ denotes the aggregated signal-to-noise ratio (SNR) at the eavesdroppers. To quantify the secrecy throughput of the system, we define the weighted secrecy sum-rate (WSSR) as
\begin{align}
{\mathcal R}=\sum_{k=1}^{K}w_k[\log_2\left(1+\gamma_k\right)-\log_2\left(1+{\overline{\gamma}}_k\right),0]^{+}\label{Secrecy_Rate_System}
\end{align}
for non-negative weights $\{w_k\}_{k=1}^{K}$ corresponding to desired quality-of-services (QoSs) of UTs.
\subsection{Problem Formulation}
Our ultimate goal is to find the system design that optimizes the secrecy throughput. This means that we strive to jointly design the beamforming matrix $\mathbf W$ and the antenna selection matrix $\bm\Delta$, such that the secrecy sum-rate term $\mathcal{R}$ is maximized. Consequently, our design problem is formulated as
\begin{equation}\label{P_1}
\begin{split}
\max_{{\mathbf W},{\bm\Delta}}~&{\mathcal R}=\sum_{k=1}^{K}w_k{\mathcal R}_{k}\\
\quad{\rm{s.t.}}~&{\mathcal C}_1:{\mathsf{tr}}({\mathbf W}{\mathbf W}^{\mathsf H})=\sum_{k=1}^{K}{\mathbf w}_k^{\mathsf H}{\mathbf w}_k\leq p,\\
&{\mathcal C}_2:s_m\in\{0,1\}~{\text{for}}~m\in[M],{\mathbf 1}^{\mathsf T}{\mathbf s}=N,
\end{split}\tag{${\mathcal{P}}_1$}
\end{equation}
where ${\mathcal C}_1$ represents the transmit power constraint. Problem \eqref{P_1} is a non-convex optimization problem due to the non-convexity of $\mathcal{R}$ with respect to $\bm\Delta$ and $\mathbf W$. In addition, considering the discrete constraints in ${\mathcal C}_2$, \eqref{P_1} is an MINLP problem, which is difficult to solve directly and to derive a globally optimal solution. In the sequel, we develop an efficient framework to approximate the optimal design via a feasible computational complexity.
\section{Proposed Solution}
By invoking the FP framework \cite{Shen2018}, we simplify problem \eqref{P_1} to a more tractable yet equivalent form. We then handle the resulting equivalent problem via the PDD \cite{Shi2020} that efficiently addresses non-convex non-smooth problems with coupling equality constraints.
\subsection{Reformulation of Problem \eqref{P_1}}
The operator $[\cdot]^{+}$ in \eqref{Secrecy_Rate_System} makes the objective of problem \eqref{P_1} intractable. To take off this operator, we convert problem \eqref{P_1} into its equivalent form as follows.
\begin{lemma}\label{Lemma_Initial_Trans}
Define ${\mathbf b}=[b_1,\ldots,b_K]^{\mathsf T}$. Problem \eqref{P_1} is equivalent to the problem defined as follows:
\begin{equation}\label{P_2}
\begin{split}
\max_{{\mathbf W},{\bm\Delta},{\mathbf b}}~&\overline{\mathcal R}=\sum_{k=1}^{K}b_k(\log_2\left(1\!+\!\gamma_k\right)-\log_2\left(1\!+\!{\overline{\gamma}}_k\right))\\
\quad{\rm{s.t.}}~&{\mathcal C}_1,{\mathcal C}_2,{\mathcal C}_3:b_k\in[0,w_k],~{\text{for}}~k\in[K].
\end{split}\tag{${\mathcal{P}}_2$}
\end{equation}
This means that the solutions $\{{\mathbf W},{\bm\Delta}\}$ for both the problems are identical.
\end{lemma}
\begin{IEEEproof}
Please refer to Appendix \ref{Proof_Lemma_Initial_Trans} for more details.
\end{IEEEproof}
We further rewrite the objective of \eqref{P_2} as
\begin{align}
\overline{\mathcal R}=\sum_{k=1}^{K}b_k\left(\log_2\left(1+\gamma_k\right)+\log_2\left(\frac{1}{1+{\overline{\gamma}}_k}\right)\right).
\end{align}
This \emph{sum-of-functions-of-ratio} form motivates us to exploit the FP framework to further simplify \eqref{P_2}. We start the derivations by the following lemma:
\begin{lemma}\label{Lemma_FP_Must}
Define $g\triangleq\sum_{j=1}^{J}\frac{p}{\delta_j^2}\lVert{\mathbf g}_j\rVert^2$. Problem \eqref{P_2} is equivalent to the problem defined as follows:
\begin{equation}\label{P_3}
\begin{split}
\max_{{\mathbf W},{\bm\Delta},{\mathbf b}}&\hat{\mathcal R}=\sum_{k=1}^{K}\!\!b_k\!\!\left(\!\log_2\left(1\!+\!\gamma_k\right)\!+\!\log_2\left(\!1\!+\!\frac{g\!-\!{\overline{\gamma}}_k}{1\!+\!{\overline{\gamma}}_k}\!\right)\!\!\right)\\
\quad{\rm{s.t.}}~&{\mathcal C}_1,{\mathcal C}_2,{\mathcal C}_3,
\end{split}\tag{${\mathcal{P}}_3$}
\end{equation}
where $g-{\overline{\gamma}}_k\geq0$.
\end{lemma}
\begin{IEEEproof}
Please refer to Appendix \ref{Proof_Lemma_FP_Must} for more details.
\end{IEEEproof}
We next exploit the FP framework to convert the fractional programming problem \eqref{P_3} equivalently to a new problem which involves a more tractable objective function.
\begin{lemma}\label{Lemma_Dual}
Problem \eqref{P_3} is equivalent to:
\begin{equation}\label{P_4}
\begin{split}
\max_{{\mathbf W},{\bm\Delta},{\mathbf b},{\bm\alpha},{\bm\beta}}&
{\mathcal F}({\mathbf W},{\bm\Delta},{\mathbf b},{\bm\alpha},{\bm\beta})=
\sum_{k=1}^{K}b_k(f_{1}^{k}+f_{2}^{k})\\
\quad{\rm{s.t.}}~&{\mathcal C}_1,{\mathcal C}_2,{\mathcal C}_3,
\end{split}\tag{${\mathcal{P}}_4$}
\end{equation}
where ${\bm\alpha}=[\alpha_1,\ldots,\alpha_K]^{\mathsf T}$, ${\bm\beta}=[\beta_1,\ldots,\beta_K]^{\mathsf T}$,
\begin{align}
&f_{1}^{k}=\log(1+\alpha_k)-\alpha_k+\frac{(1+\alpha_k)\lvert{\mathbf h}_k^{\mathsf H}{\bm\Delta}{\mathbf w}_k\rvert^2}{\sum_{i=1}^{K}\lvert{\mathbf h}_k^{\mathsf H}{\bm\Delta}{\mathbf w}_{i}\rvert^2+\sigma_k^2},\\
&f_{2}^{k}=\log(1+\beta_k)-\beta_k+{(1+\beta_k)(g-{\overline{\gamma}}_k)}
({1+g})^{-1}.
\end{align}
The optimal $\alpha_k$ and $\beta_k$ are given by $\alpha_k^{\star}=\gamma_k$ and $\beta_k^{\star}=\frac{g-{\overline{\gamma}}_k}{1+{\overline{\gamma}}_k}$, respectively.
\end{lemma}
\begin{IEEEproof}
Please refer to Appendix \ref{Proof_Lemma_Dual} for more details.
\end{IEEEproof}
It is worth noting that $g$ is independent of $\{{\mathbf W},{\bm\Delta},{\mathbf b}\}$. Therefore, the intractability of ${\mathcal F}({\mathbf W},{\bm\Delta},{\mathbf b},{\bm\alpha},{\bm\beta})$ mainly originates from the fractional terms involved in $\{f_{1}^{k}\}_{k=1}^{K}$. To handle this difficulty, we introduce auxiliary variables ${\bm\eta}=[\eta_1,\ldots,\eta_K]^{\mathsf T}$ to simplify ${\mathcal F}({\mathbf W},{\bm\Delta},{\mathbf b},{\bm\alpha},{\bm\beta})$. Then, the following lemma can be found.
\begin{lemma}\label{Lemma_Quad}
Problem \eqref{P_4} is equivalent to:
\begin{equation}\label{P_5}
\begin{split}
\max_{{\mathbf W},{\bm\Delta},{\mathbf b},{\bm\alpha},{\bm\beta},{\bm\eta}}&
{\mathcal L}({\mathbf W},{\bm\Delta},{\mathbf b},{\bm\alpha},{\bm\beta},{\bm\eta})\!=\!
\sum_{k=1}^{K}\!b_k(g_{1}^{k}\!+\!f_{2}^{k})\\
\quad{\rm{s.t.}}~&{\mathcal C}_1,{\mathcal C}_2,{\mathcal C}_3,
\end{split}\tag{${\mathcal{P}}_5$}
\end{equation}
where $g_{1,k}=\log(1+\alpha_k)-\alpha_k+(1+\alpha_k){\overline{g}}_1^{k}$ with
\begin{equation}
\begin{split}
{\overline{g}}_1^{k}=2\Re\{{\eta}_k^{*}{\mathbf h}_k^{\mathsf H}{\bm\Delta}{\mathbf w}_k\}
\!-\!|\eta_k|^2\big(\sum_{i=1}^{K}\!|{\mathbf h}_k^{\mathsf H}{\bm\Delta}{\mathbf w}_{i}|^2\!+\!\sigma_k^2\big).
\end{split}
\end{equation}
The optimal $\eta_k$ is given by $\eta_k^{\star}=(\sum_{i=1}^{K}\!|{\mathbf h}_k^{\mathsf H}{\bm\Delta}{\mathbf w}_{i}|^2\!+\!\sigma_k^2)^{-1}{\mathbf h}_k^{\mathsf H}{\bm\Delta}{\mathbf w}_k$.
\end{lemma}
\begin{IEEEproof}
Similar to the proof of Lemma \ref{Lemma_Dual}.
\end{IEEEproof}
\begin{remark}
Lemmas \ref{Lemma_Initial_Trans} to \ref{Lemma_Quad} indicate that \eqref{P_5} is a variational form of the principal problem \eqref{P_1}. This means that the solutions $\{{\mathbf W},{\bm\Delta}\}$ for both the problems are identical. It is worth noting that unlike the original form in problem \eqref{P_1}, the problem \eqref{P_5} has an objective function that is marginally convex over each of the variable ${\mathbf W}$, ${\bm\Delta}$, ${\mathbf b}$, ${\bm\alpha}$, ${\bm\beta}$, and ${\bm\eta}$.
\end{remark}
Furthermore, to tackle the discrete constraints in ${\mathcal C}_2$, we define the auxiliary variables $\overline{\mathbf{s}}=[\overline{s}_1,\ldots,\overline{s}_M]^{\mathsf T}$ which satisfy the following constraints: ${\overline s}_m=s_m$ and $s_m\left(1-{\overline s}_m\right)=0$. We thus can equivalently find the solution of \eqref{P_1} by solving the following optimization:
\begin{equation}\label{P_6}
\begin{split}
&\max_{{\mathbf W},{\bm\Delta},{\mathbf b},{\bm\alpha},{\bm\beta},{\bm\eta},\overline{\mathbf s}}
{\mathcal L}({\mathbf W},{\bm\Delta},{\mathbf b},{\bm\alpha},{\bm\beta},{\bm\eta})\\
&\quad{\rm{s.t.}}~{\mathcal C}_1,{\mathcal C}_2,{\mathcal C}_3,\\
&\qquad~~{\mathcal C}_4:{\overline s}_m=s_m,s_m\left(1-{\overline s}_m\right)=0,~{\text{for}}~m\in[M].
\end{split}\tag{${\mathcal{P}}_6$}
\end{equation}
We note that in \eqref{P_6}, variables $\mathbf s$ and $\overline{\mathbf s}$ are only constrained through ${\overline s}_m=s_m$, $s_m\left(1-{\overline s}_m\right)=0$, and ${\mathbf 1}^{\mathsf T}{\mathbf s}=N$. To handle these equality constraints, we resort to the PDD technique.
\vspace{-5pt}
\subsection{The Proposed PDD-Based Algorithm}
The PDD-based algorithm is characterized by an embedded double loop structure \cite{Shi2020}. The inner loop solves the augmented Lagrangian (AL) subproblem while the outer loop updates the dual variables and the penalty parameters that correspond to constraint violation. We obtain the AL problem by moving the equality constraints as a penalty term to the objective function. By \cite{Shi2020}, the AL problem corresponding to \eqref{P_6} is given by
\begin{equation}\label{P_7}
\begin{split}
\max_{{\mathbf W},{\bm\Delta},{\mathbf b},{\bm\alpha},{\bm\beta},{\bm\eta},\overline{\mathbf s}}
{\mathcal L}({\mathbf W},{\bm\Delta},{\mathbf b},{\bm\alpha},{\bm\beta},{\bm\eta})-f_{\rho}
\quad{\rm{s.t.}}~{\mathcal C}_1,{\mathcal C}_3.
\end{split}\tag{${\mathcal{P}}_7$}
\end{equation}
where $\rho>0$ is the penalty parameter penalizing the violation of the equality constraints, and $f_\rho$ is given by
\begin{align}
f_{\rho}&=\frac{1}{2\rho}\left[\left({\mathbf 1}^{\mathsf T}{\mathbf s}\!-\!N\!+\!\rho\xi\right)^2
+\sum_{m=1}^{M}\left[\left(s_m\!-\!{\overline s}_m\!+\!\rho\mu_m\right)^2\right.\right.\nonumber\\
&\left.\left.+\left(s_m\left(1-{\overline s}_m\right)+\rho\lambda_m\right)^2\right]\right],
\end{align}
with $\xi$, ${\bm\mu}=[\mu_1,\ldots,\mu_M]^{\mathsf T}$, and ${\bm\lambda}=[\lambda_1,\ldots,\lambda_M]^{\mathsf T}$ denoting the Lagrangian dual variables associated with the equality constraints in ${\mathbf 1}^{\mathsf T}{\mathbf s}=T$, ${\overline s}_m=s_m$, and $s_m\left(1-{\overline s}_m\right)=0$, respectively. It is observed that as $\rho\rightarrow0$, the penalty term is forced to zero, i.e., equality constraints are enforced. It is shown in \cite{Shi2020} that updating the primal and dual variables, as well as the penalty factor in an alternating manner, PDD converges to a stationary-point solution. Thus, in the following, we focus on solving problem \eqref{P_7} by invoking the block coordinate descent (BCD) method. More specifically, we sequentially update ${\mathbf W}$, $\bm\Delta$ (or ${\mathbf s}$), $\overline{\mathbf s}$, ${\mathbf b}$, $\bm\alpha$, $\bm\beta$, and $\bm\eta$ in the inner loop through the following marginal optimizations.
\subsubsection{Optimizing $\mathbf W$}
The marginal problem for $\mathbf{W}$ reads
\begin{equation}\label{P_8}
\begin{split}
\min_{\mathbf W}~
\sum_{k=1}^{K}({\mathbf w}_k^{\mathsf H}{\mathbf A}_k{\mathbf w}_k-2\Re\{{\mathbf w}_k^{\mathsf H}{\mathbf a}_k\})\quad{\rm{s.t.}}~{\mathcal C}_1.
\end{split}\tag{${\mathcal{P}}_8$}
\end{equation}
where ${\mathbf a}_k=b_k(1+\alpha_k){\eta}_k{\bm\Delta}^{\mathsf H}{\mathbf h}_k$ and
\begin{equation}
\begin{split}
{\mathbf A}_k&=\sum_{i=1}^{K}b_i(1+\alpha_i)|\eta_i|^2{\bm\Delta}^{\mathsf H}{\mathbf h}_i{\mathbf h}_i^{\mathsf H}{\bm\Delta}\\
&+{b_k(1+\beta_k)}{({1+g})^{-1}}\sum_{j=1}^{J}{\bm\Delta}^{\mathsf H}{\mathbf g}_j{\mathbf g}_j^{\mathsf H}{\bm\Delta}\succeq{\mathbf 0}.
\end{split}
\end{equation}
This is a standard convex quadratic optimization subproblem whose solution is given by
\begin{align}
{\mathbf w}_k^{\star}=({\mathbf{A}}_k+\lambda{\mathbf I})^{-1}{\mathbf{a}}_k,~{\text{for}}~k\in[K].
\end{align}
The regularizer $\lambda$ is chosen, such that the complementarity slackness condition, i.e., $\lambda({\mathsf{tr}}({\mathbf{W}}{\mathbf{W}}^{\mathsf H})- p)=0$, is satisfied. If $\sum_{k=1}^{K}{\mathbf a}_k^{\mathsf H}{\mathbf A}_k^{-2}{\mathbf a}_k=p$; then, $\lambda=0$. Otherwise, we can obtain the solution of $\lambda$ from the following identity:
\begin{align}
{\mathsf{tr}}({\mathbf{W}}{\mathbf{W}}^{\mathsf H})=\sum_{k=1}^{K}{\mathbf a}_k^{\mathsf H}({\mathbf{A}}_k+\lambda{\mathbf I})^{-2}{\mathbf a}_k=p.
\end{align}
It follows that
\begin{align}
\sum_{k=1}^{K}\sum_{m=1}^{M}\frac{|[{\mathbf U}_{{\mathbf A}_k}{\mathbf a}_k]_{m}|^2}{([{\bm\Lambda}_{{\mathbf A}_k}]_{m,m}+\lambda)^{2}}=p,\label{Power_Bisection}
\end{align}
where ${\mathbf A}_k={\mathbf U}_{{\mathbf A}_k}^{\mathsf H}{\bm\Lambda}_{{\mathbf A}_k}{\mathbf U}_{{\mathbf A}_k}$ is the eigen-decomposition of ${\mathbf A}_k$. Since $[{\bm\Lambda}_{{\mathbf A}_k}]_{m,m}\geq0$ for $m\in[M]$, the left-hand side of \eqref{Power_Bisection} is a monotonously decreasing function of $\lambda\geq0$. Consequently, we can find $\lambda$ by solving equation \eqref{Power_Bisection} using the bisection-based search method.
\subsubsection{Optimizing $\mathbf s$}
Define ${\mathbf W}_k={\mathsf{diag}}\{{\mathbf w}_k\}$ for $k\in[K]$,
\begin{align}
{\mathbf Q}&\triangleq{\Re}\bigg\{
\sum_{i=1}^{K}\sum_{k=1}^{K}w_k(1+\alpha_k)|\eta_k|^2{\mathbf W}_i^{\mathsf H}{\mathbf h}_k{\mathbf h}_k^{\mathsf H}{\mathbf W}_i\nonumber\\
&+
\sum_{k=1}^{K}\frac{b_k(1+\beta_k)}{1+g}\sum_{j=1}^{J}{\mathbf W}_k^{\mathsf H}{\mathbf g}_j{\mathbf g}_j^{\mathsf H}{\mathbf W}_k
+\frac{1}{2\rho}\nonumber\\
&\times({\mathbf 1}{\mathbf 1}^{\mathsf{T}}+{\mathbf I}+{{\mathsf{diag}}\{\left(1-{\overline s}_1\right)^2,\cdots,\left(1-{\overline s}_M\right)^2\}})\big\},
\end{align}
\begin{align}
{\mathbf g}&\triangleq2\Re\!\left\{{\mathbf q}\right\}\!-\!\frac{1}{\rho}\left(\left(\rho\xi\!-\!N\right){\mathbf 1}\!+\!\left(\rho{\bm\mu}\!-\!\overline{\mathbf s}\right)\!+\!\rho\left({\mathbf{1}}\!-\!\overline{\mathbf s}\right)\!\odot\!{\bm\lambda}\right),\\
{\mathbf q}&\triangleq\sum_{k=1}^{K}(1+\alpha_k)\Re\{{\eta}_k{\mathbf W}_k^{\mathsf H}{\mathbf h}_k\},
\end{align}
The marginal problem for $\mathbf s$ is given by
\begin{equation}\label{Antenna_Selection_Matrix_Opt}
{\mathbf s}^{\star}=\argmin_{{\mathbf s}\in{\mathbbmss R}^{M}}\left({\mathbf s}^{\mathsf T}{\mathbf Q}{\mathbf s}-{\mathbf s}^{\mathsf T}{\mathbf g}\right).
\end{equation}
It is readily shown that ${\mathbf Q}\succ{\mathbf 0}$, which means that the problem \eqref{Antenna_Selection_Matrix_Opt} is a standard convex problem. Using its first-order optimality condition, we obtain ${{\mathbf s}}^{\star}=\frac{1}{2}{\mathbf Q}^{-1}{\mathbf g}$.
\subsubsection{Optimizing $\overline{\mathbf s}$}
The marginal optimization for $\{\overline{s}_m\}_{m=1}^{M}$ decouples into $M$ scalar optimization problems with solution to the $m$th problem being
\begin{equation}\label{Problem_Auxiliary}
\overline{s}_m^{\star}=\argmin_{\overline{s}_m\in{\mathbbmss R}}\left(\overline{a}_m\overline{s}_m^2-2\overline{b}_m\overline{s}_m\right).
\end{equation}
Here, $\overline{a}_m=1+s_m^2$ and $\overline{b}_m=s_m+\rho\mu_m+s_m^2+s_m\rho\lambda_m$. The optimization in \eqref{Problem_Auxiliary} is a quadratic scalar problem whose solution is given by ${\overline{s}}_m^{\star}={\overline{b}_m}/{\overline{a}_m}$.
\subsubsection{Optimizing ${\mathbf b}$}
The marginal optimization for $\{b_k\}_{k=1}^{K}$ decouples into $K$ scalar optimization problems with solution to the $k$th problem being
\begin{equation}\label{Problem_Weights}
\begin{split}
b_k^{\star}&=\argmax_{b_k\in[0,w_k]}\left(b_k(g_{1}^{k}+f_{2}^{k})\right)\\
&=w_k{\mathbf 1}_{\{g_1^{k}>-f_2^{k}\}},
\end{split}
\end{equation}
where ${\mathbf 1}_{\{\cdot\}}$ represents the indicator function.
\subsubsection{Optimizing the Auxiliary Variables}
For given $\{{\mathbf W},{\mathbf s},\overline{\mathbf s},{\mathbf b}\}$, the auxiliary variables $\bm\alpha$, $\bm\beta$, and $\bm\eta$ can be updated by Lemmas \ref{Lemma_Dual} and \ref{Lemma_Quad}. The marginally optimal $\bm\alpha$, $\bm\beta$, and $\bm\eta$ are given by $\alpha_k^{\star}=\gamma_k$, $\beta_k^{\star}=\frac{g-{\overline{\gamma}}_k}{1+{\overline{\gamma}}_k}$, and $\eta_k^{\star}=(\sum_{i=1}^{K}\!|{\mathbf h}_k^{\mathsf H}{\bm\Delta}{\mathbf w}_{i}|^2\!+\!\sigma_k^2)^{-1}{\mathbf h}_k^{\mathsf H}{\bm\Delta}{\mathbf w}_k$, respectively.

\begin{algorithm}[!t]
\algsetup{linenosize=\tiny} \scriptsize
  \caption{PDD-based algorithm for solving problem \eqref{P_1}}
  \label{Algorithm1}
  \begin{algorithmic}[1]
    \STATE Initialize primary variables $\left\{{\mathbf W},{\mathbf s},\overline{\mathbf s},{\mathbf b},{\bm\alpha},{\bm\beta},{\bm\eta}\right\}$, dual variables $\left\{\xi,{\bm\lambda},{\bm\mu}\right\}$, threshold $\mu$, penalty factor $\rho>0$, and scaling factor $\chi\in(0,1)$
    \REPEAT
    \REPEAT
      \STATE Update $\left\{{\mathbf W},{\mathbf s},\overline{\mathbf s},{\mathbf b},{\bm\alpha},{\bm\beta},{\bm\eta}\right\}$ by the BCD method
    \UNTIL{convergence}
    \IF{$h < \mu$}
	\STATE Set $\xi=\xi+{\rho}^{-1}({\mathbf 1}^{\mathsf T}{\mathbf s}-N)$, ${\bm\mu}={\bm\mu}+{\rho}^{-1}\left({\overline{\mathbf s}}-{\mathbf s}_n\right)$, and ${\bm\lambda}={\bm\lambda}+{\rho}^{-1}{\mathbf s}\odot\left(1-{\overline{\mathbf s}}\right)$
	\ELSE
	\STATE Set $\rho = \chi\rho$
	\ENDIF
    \STATE Set $\mu=\chi h$ and $t=t+1$
    \UNTIL{convergence}
  \end{algorithmic}
\end{algorithm}

Finally, the dual variables $\{{\xi},{\bm\lambda},{\bm\mu}\}$ and penalty factor $\rho$ can be updated following the PDD framework. As a consequence, the overall algorithm for solving problem \eqref{P_1} is summarized in Algorithm \ref{Algorithm1}, where $h\triangleq\max_{\forall m}\{|{\mathbf 1}\!^{\mathsf T}{\mathbf s}\!-\!N|,\left|{\overline s}_m\!-\!s_m\right|,\left|s_m\!\left(1\!-\!{\overline s}_m\right)\right|\}$ denotes the constraint violation function. The computational complexity of the proposed algorithm can be further characterized in terms of problem dimensions. To this end, let $I_{\rm{out}}$ and $I_{\rm{fp}}$ denote the numbers of iterations in the outer loop and the inner-loop FP, respectively. It is readily shown that the complexity of marginal optimizations with respect to ${\mathbf W}$, ${\mathbf s}$, $\overline{\mathbf s}$, ${\mathbf b}$, $\bm\alpha$, $\bm\beta$, and $\bm\eta$ scale with ${\mathcal O}\left(M^3\right)$, ${\mathcal O}\left(M^3\right)$, ${\mathcal O}\left(M\right)$, ${\mathcal O}\left(KM^2\right)$, ${\mathcal O}\left(K\right)$, ${\mathcal O}\left(K\right)$, and ${\mathcal O}\left(K\right)$, respectively. Therefore, the overall complexity of Algorithm \ref{Algorithm1} scales with ${\mathcal O}(I_{\text{out}}I_{\text{fp}}(2M^3+KM^2))$.

\begin{figure}[!t]
    \centering
    \subfigure[Convergence of the WSSR.]
    {
        \includegraphics[height=0.25\textwidth]{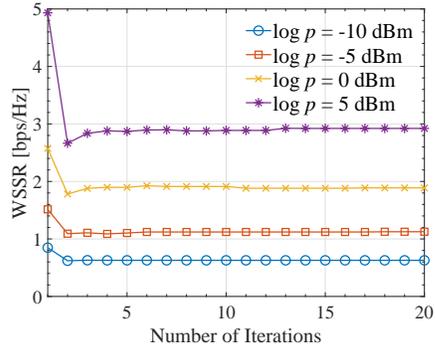}
	   \label{fig1a}	
    }
   \subfigure[Constraint violation.]
    {
        \includegraphics[height=0.25\textwidth]{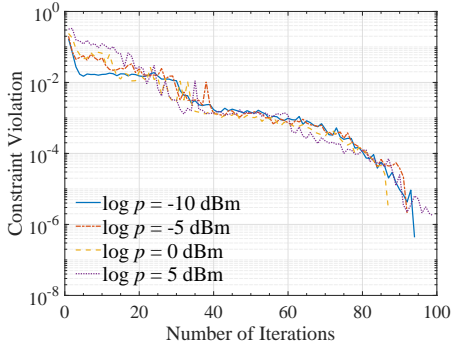}
	   \label{fig1b}	
    }
\caption{Average convergence performances for $L=6$.}
    \label{figure1}
\end{figure}

\subsection{Low-Complexity Sequential Optimization}
Although the PDD-based algorithm leads to a feasible complexity, it can lead to a computational burden in many applications. To this end, we develop an alternative scheme based on sequential optimization approach. In this respect, we first design the beamforming matrix using the FP-based method; then, we select the antennas using a greedy search (GS). These two steps are illustrated in the sequel.
\subsubsection{Transmit Precoding Design}
At the first step of the SO-based method, we design the beamforming matrix $\mathbf W$ by setting ${\bm\Delta}={\mathbf I}$. The resulting problem can be solved by using the FP-based method. The detailed steps are similar to those outlined in the previous part and are omitted here for brevity.
\subsubsection{Antenna Selection}
In the next step, we optimize the antenna selection vector $\mathbf s$ for the updated beamformer. The resulting problem of antenna selection is given by
\begin{align}\label{SO_2}
{\mathbf s}^{\star}=\argmax{\mathcal R},~{\rm{s.t.}}~{\mathcal C}_2.
\end{align}
To tackle the discrete constraint, we exploit the GS method. More details about this method can be found in \cite{Gershman2004} and are omitted here due to page limitations.

Based on \cite{Gershman2004}, we can show that the complexity of sequential optimizations with respect to ${\mathbf s}$ and $\mathbf W$ scale with ${\mathcal O}(KNM)$, and ${\mathcal O}(I_{\text{fp}}M^3)$, respectively. Hence, the complexity of the SO-based method scales with ${\mathcal O}(KNM+I_{\text{fp}}M^3)$, which is lower than the PDD-based method.

\begin{figure}[!t]
    \centering
    \subfigure[WSSR vs. the power budget.]
    {
        \includegraphics[height=0.25\textwidth]{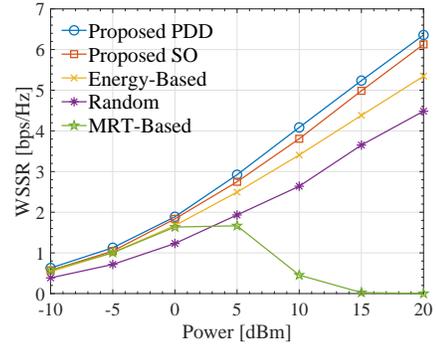}
	   \label{fig2a}	
    }
   \subfigure[WSSR vs. the RF chain number.]
    {
        \includegraphics[height=0.25\textwidth]{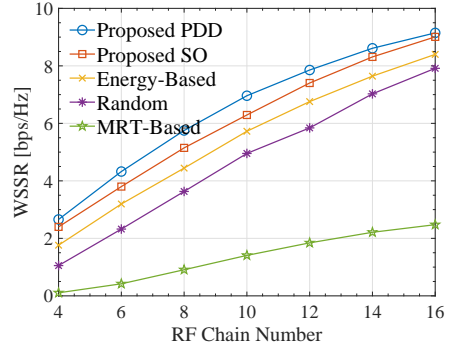}
	   \label{fig2b}	
    }
\caption{WSSR performance: (a) $N=6$, (b) $\log{p}=10$ dBm.}
    \label{figure2}
\end{figure}

\section{Numerical Results}
In this section, simulation results are provided to verify the effectiveness of the proposed algorithms. For simulations, the following parameters are used unless explicitly mentioned otherwise: $M=24$, $K=6$, $J=4$, $w_k=1$, $\log{\sigma_k^2}=\log{\delta_j^2}=-120$ dBm, $\mu=1$, $\rho=1$, and $\chi=0.1$. We further generate the channel realizations as follows. Regarding the large-scale fading, we assume that all the channels, i.e., $\{{\mathbf h}_k\}_{k=1}^{K}$ and $\{{\mathbf g}_j\}_{j=1}^{J}$, exhibit the same path loss $-120$ dB for illustration. Meanwhile, for the small-scale fading of the channels, we consider the standard Rayleigh fading model. All the optimization variables are randomly initialized, and all simulation curves are averaged over $500$ independent channel realizations.

In {\figurename} {\ref{figure1}}, we first study the convergence behavior of the proposed PDD-based method. It can be seen from {\figurename} {\ref{fig1a}} that the WSSR rapidly converges to a stationary value. {\figurename} {\ref{fig1b}} shows the constraint violation in terms of the number of outer iterations. We observe that the constraint violation reduces to a threshold $10^{-4}$ in less than $100$ outer iterations, meaning that the solution has essentially met the equality constraints for problem \eqref{P_1}.

We next consider the following benchmark schemes for performance comparison: 1) Random scheme, in which $\mathbf s$ is randomly set, and ${\mathbf W}$ is optimized by the FP-based method. 2) Energy-based scheme, in which the BS antennas corresponding to the $N$ strongest channel gains are selected, and ${\mathbf W}$ is optimized by the FP-based method. This selection strategy was utilized in previous works \cite{Asaad2018_JSAC,Ouyang2020}. 3) MRT-based scheme, in which $\mathbf W$ is generated by the MRT protocol, and $\mathbf s$ is designed by the GS-based method \cite{Bereyhi2021,Bereyhi2018}.

{\figurename} {\ref{figure2}} compares the WSSR performances achieved by the proposed methods and the benchmark methods. In {\figurename} {\ref{fig2a}}, the WSSR is plotted as a function of the transmit power budget $p$ for different optimization schemes assuming $N=6$. From this graph, we find that the proposed PDD-based method has the best secrecy performance, followed by the SO-based scheme. As stated before, the SO-based method involves less computational complexity than the PDD-based scheme. The above results imply that the SO-based method is preferred for systems with highly restricted computational capacity. What stands out in this graph is that the methods using FP-based secure beamforming, i.e., the PDD-based, SO-based, Energy-based, and Random schemes, are far superior to the MRT-based scheme in terms of the WSSR, especially in the high-SNR regime. This observation highlights the superiority of the proposed FP-based beamforming design.

We next fix $\log{p}=10$ dBm and plot the WSSR versus the number of RF chains for selected BS antennas. The results are shown in {\figurename} {\ref{fig2b}} implying that in all the presented cases, the increase in the number of RF chains improves the secrecy performance. Specifically, the proposed two methods significantly outperform the benchmarks in terms of the WSSR. It validates the advantages of using our proposed joint design to enhance secure transmissions at the PHY.
\section{Conclusion}
We proposed an iterative algorithm capitalizing on the FP and PDD techniques for joint antenna selection and transmit beamforming in MIMO wiretap channels. We also developed an alternative algorithm with reduced computational complexity based on the SO scheme. Our numerical results imply that the PDD-based method outperforms the SO-based and benchmark methods. The SO-based algorithm closely tracks the performance of the PDD-based method while benefiting considerably in terms of complexity.
\begin{appendix}
\subsubsection{Proof of Lemma \ref{Lemma_Initial_Trans}}\label{Proof_Lemma_Initial_Trans}
We prove this lemma by showing that \eqref{P_2} can be equivalently transformed to \eqref{P_1}. Given $\{{\mathbf W},{\bm\Delta}\}$, the optimal $b_k$ satisfies $b_k^{\star}=w_k{\mathbf 1}_{\{\gamma_k>{\overline{\gamma}}_k\}}$. We note that substituting $b_k=b_k^{\star}$ into $\overline{\mathcal R}$ recovers the objective of \eqref{P_1}:
\begin{align}
\overline{\mathcal R}
&=\sum_{k=1}^{K}{\mathbf 1}_{\{\gamma_k>{\overline{\gamma}}_k\}}w_k(\log_2\left(1\!+\!\gamma_k\right)\!-\!\log_2\left(1\!+\!{\overline{\gamma}}_k\right))\\
&=\sum_{k=1}^{K}w_k[\log_2\left(1\!+\!\gamma_k\right)\!-\!\log_2\left(1\!+\!{\overline{\gamma}}_k\right),0]^{+}={\mathcal R}.
\end{align}
The final results follow immediately.
\subsubsection{Proof of Lemma \ref{Lemma_FP_Must}}\label{Proof_Lemma_FP_Must}
By the Cauchy inequality, we have
\begin{align}
{\overline{\gamma}}_k&\leq
\sum_{j=1}^{J}{\delta_j^{-2}}\lVert{\mathbf g}_j\rVert^2\lVert{\bm\Delta}{\mathbf w}_k\rVert^2.
\end{align}
We next exploit the following fact:
\begin{align}
\lVert{\bm\Delta}{\mathbf w}_k\rVert^2=
{\mathbf w}_k^{\mathsf H}{\bm\Delta}{\mathbf w}_k\leq{\mathbf w}_k^{\mathsf H}{\mathbf I}{\mathbf w}_k=\lVert{\mathbf w}_k\rVert^2,
\end{align}
to obtain the following relationship:
\begin{align}
{\overline{\gamma}}_k&\leq
\sum_{j=1}^{J}{\delta_j^{-2}}\lVert{\mathbf g}_j\rVert^2\lVert{\mathbf w}_k\rVert^2=g.
\end{align}
It follows that
\begin{align}
\log_2\left(\frac{1}{1+{\overline{\gamma}}_k}\right)&=\log_2\left(\frac{1+g}{1+{\overline{\gamma}}_k}\right)-\log_2\left(1+g\right)\\
&=\log_2\left(1+\frac{g-{\overline{\gamma}}_k}{1+{\overline{\gamma}}_k}\right)-\log_2\left(1+g\right).
\end{align}
Note that $\log_2\left(1+g\right)$ is independent of $\{{\mathbf W},{\bm\Delta}\}$. Hence, we can establish the equivalence between \eqref{P_2} and \eqref{P_3}.
\subsubsection{Proof of Lemma \ref{Lemma_Dual}}\label{Proof_Lemma_Dual}
Since $\mathcal F$ is concave over $\bm\alpha$ and $\bm\beta$ for fixed $\{{\mathbf W},{\bm\Delta},{\mathbf b}\}$, we take its complex derivative and solve each $\frac{\partial}{\partial \alpha_k}{\mathcal F}=0$ and $\frac{\partial}{\partial \beta_k}{\mathcal F}=0$. The optimal $\alpha_k$ and $\beta_k$ are easily seen as $\alpha_k^{\star}$ and $\beta_k^{\star}$, respectively. Inserting $\alpha_k^{\star}$ and $\beta_k^{\star}$ back to $\mathcal F$ recovers the objective function in \eqref{P_3}, thus establishing the equivalence of these two problems.
\end{appendix}


\begin{thebibliography}{00}
\bibitem{Chen2017} X. Chen, D. W. K. Ng, W. Gerstacker, and H.-H. Chen, ``A survey on multiple-antenna techniques for physical layer security,'' \emph{IEEE Commun. Surveys Tuts.}, vol. 19, no. 2, pp. 1027--1053, Second Quarter, 2017.
\bibitem{Ouyang2020} C. Ouyang \emph{et al.}, ``Security enhancement via antenna selection in MIMOME channels with discrete inputs,'' \emph{IEEE Trans. Commun.}, vol. 68, no. 8, pp. 5041--5055, Aug. 2020.
\bibitem{Asaad2018_JSAC} S. Asaad \emph{et al.}, ``Optimal transmit antenna selection for massive MIMO wiretap channels,'' \emph{IEEE J. Sel. Areas Commun.}, vol. 36, no. 4, pp. 817--828, Apr. 2018.
\bibitem{Tian2020} M. Tian \emph{et al.}, ``Joint beamforming design and receive antenna selection for large-scale MIMO wiretap channels,'' \emph{IEEE Trans. Veh. Technol.}, vol. 69, no. 3, pp. 2716--2730, Mar. 2020.
\bibitem{Ouyang_2019} C. Ouyang, Z. Ou, L. Zhang, and H. Yang, ``Optimal transmit antenna selection algorithm in massive MIMOME channels,'' in \emph{Proc. IEEE Wireless Commun. Netw. Conf. (WCNC)}, pp. 1--6, 2019.
\bibitem{Bereyhi2021} A. Bereyhi \emph{et al.}, ``Securing massive MIMO systems: Secrecy for free with low-complexity architectures,'' \emph{IEEE Trans. Wireless Commun.}, vol. 20, no. 9, pp. 5831--5845, Sep. 2021.
\bibitem{Bereyhi2018} A. Bereyhi \emph{et al.}, ``Iterative antenna selection for secrecy enhancement in massive MIMO wiretap channels,'' in \emph{Proc. IEEE Int. Wkshp. Signal Process. Adv. Wireless Commun. (SPAWC)}, pp. 1--5, 2018.
\bibitem{Shen2018} K. Shen and W. Yu, ``Fractional programming for communication systems--Part \uppercase\expandafter{\romannumeral1}: Power control and beamforming,'' \emph{IEEE Trans. Signal Process.}, vol. 66, no. 10, pp. 2616--2630, May. 2018.
\bibitem{Shi2020} Q. Shi and M. Hong, ``Penalty dual decomposition method for nonsmooth nonconvex optimization--Part \uppercase\expandafter{\romannumeral1}: Algorithms and convergence analysis,'' \emph{IEEE Trans. Signal Process.}, vol. 68, pp. 4108--4122, 2020.
\bibitem{Gershman2004} M. Gharavi-Alkhansari and A. B. Gershman, ``Fast antenna subset selection in MIMO systems,'' \emph{IEEE Trans. Signal Process.}, vol. 52, no. 2, pp. 339--347, Feb. 2004.
\end{thebibliography}
\end{document}